\documentclass{moriond}

\def\Journal#1#2#3#4{{#1} {\bf #2}, #3 (#4)}

\def\PLB{{\em Phys. Lett.}  B}

\def\JPD{{\em J. Phys.} D}
\def\JHEP{{\em JHEP}}

\def\be{\begin{equation}}
  \def\ee{\end{equation}}
\def\bea{\begin{eqnarray}}
           \def\eea{\end{eqnarray}}
         
         \newcommand{\MHT}{H_T^{\rm miss}}
         \newcommand{\HT}{H_T}

         \newcommand{\Photo}{}

\begin{document}
         \vspace*{4cm}
         \title{Search for supersymmetry in the multijet and missing transverse
           momentum final state: an analysis performed on 2.3 fb$^{-1}$ of 13
           TeV pp collision data collected with the CMS detector}

         \author{Jack Bradmiller-Feld for the CMS Collaboration}

         \address{Department of Physics, University of California, Santa Barbara, 93106, USA}

         \maketitle\abstracts{
           We present results from a generic search for strongly produced
           supersymmetric particles in pp collisions in the multijet + missing
           transverse momentum final state. The data sample corresponds to 2.3
           fb$^{-1}$ recorded by the CMS experiment at $\sqrt{s}=13$ TeV. This search
           is motivated by supersymmetry (SUSY) models that avoid
           fine-tuning of the higgs mass. In such models, certain strongly produced SUSY particles,
           including the gluino and top squark, are predicted to have masses on
           the order of a TeV. These particles also have some of the highest
           production cross sections in SUSY and give rise to final states with
           distinct, high jet multiplicity event signatures. To make the analysis
           sensitive to a wide range of such final states, events are classified
           by the number of jets, the scalar sum of the transverse momenta of the
           jets, the vector sum of the transverse momenta of the jets, and the
           number of b-tagged jets. No significant
           excess is observed beyond the standard model (SM) expectation. The results are
           interpreted as limits on simplified SUSY models. In these models,
           gluino masses as high as 1600 GeV are excluded at 95\% CL for
           scenarios with low $\tilde{\chi}_{1}^{0}$ mass, exceeding the most stringent limits
           set in by CMS at $\sqrt{s}=8$ TeV by more than 200 GeV in several simplified models.}

         \section{Introduction}

         Strongly produced supersymmetric particles, including gluinos and
         third generation squarks, are among the most
         attractive search targets in early $\sqrt{s}=13$ TeV LHC data. At the mass limits set on simplified models~\cite{SMS} with low $\tilde{\chi}_{1}^{0}$
         mass at $\sqrt{s}=8$ TeV, the cross section for gluino pair production, for example, is enhanced by
         a factor of thirty with respect to $\sqrt{s}=8$ TeV due to the mass-dependent
         increase in parton luminosity at $\sqrt{s}=13$ TeV~\cite{partonlumi}. Consequently, a search for strongly produced
         SUSY performed at $\sqrt{s}=13$ TeV needs only a
         fraction of the data taken at $\sqrt{s}=8$ TeV to exceed the
         sensitivity obtained in Run I for many models.

\begin{figure}
           \centering
           \includegraphics[width=0.2425\linewidth]{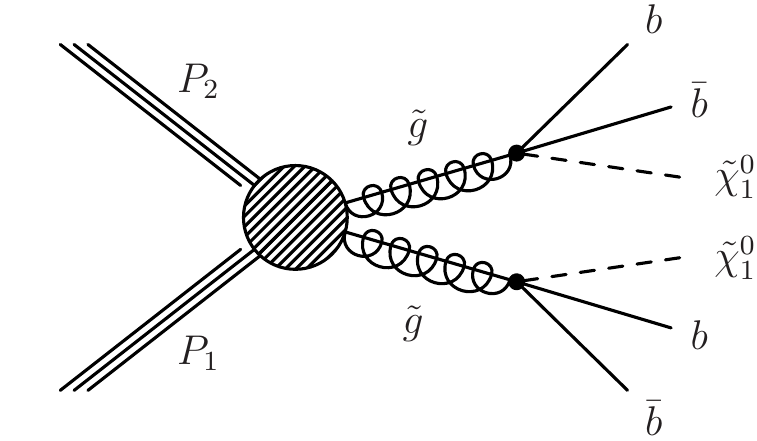}
           \includegraphics[width=0.2425\linewidth]{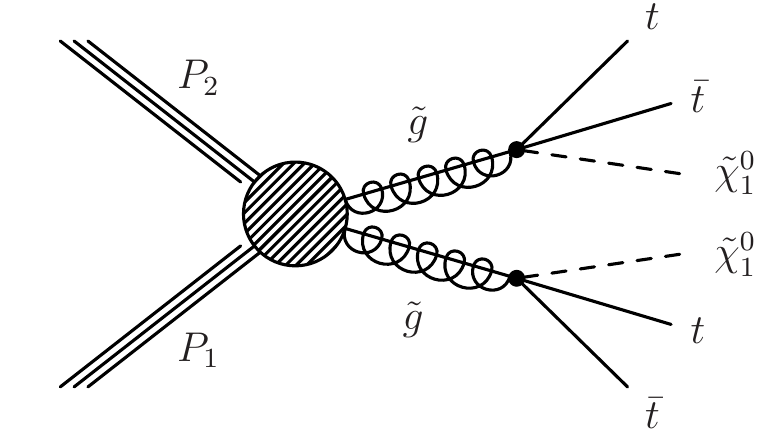}
           \includegraphics[width=0.2425\linewidth]{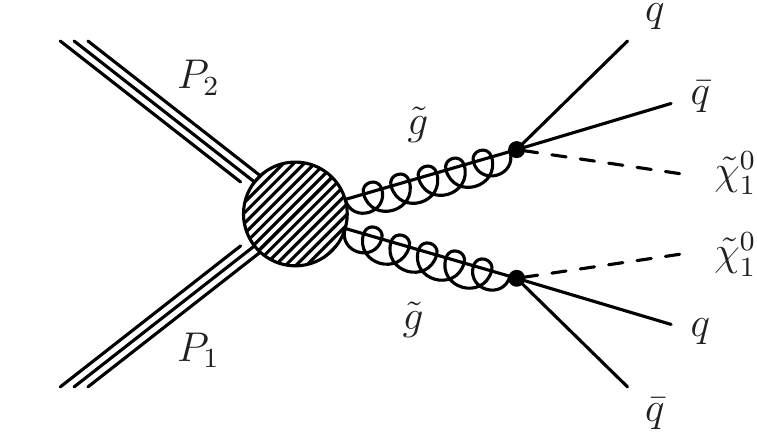}
           \includegraphics[width=0.2425\linewidth]{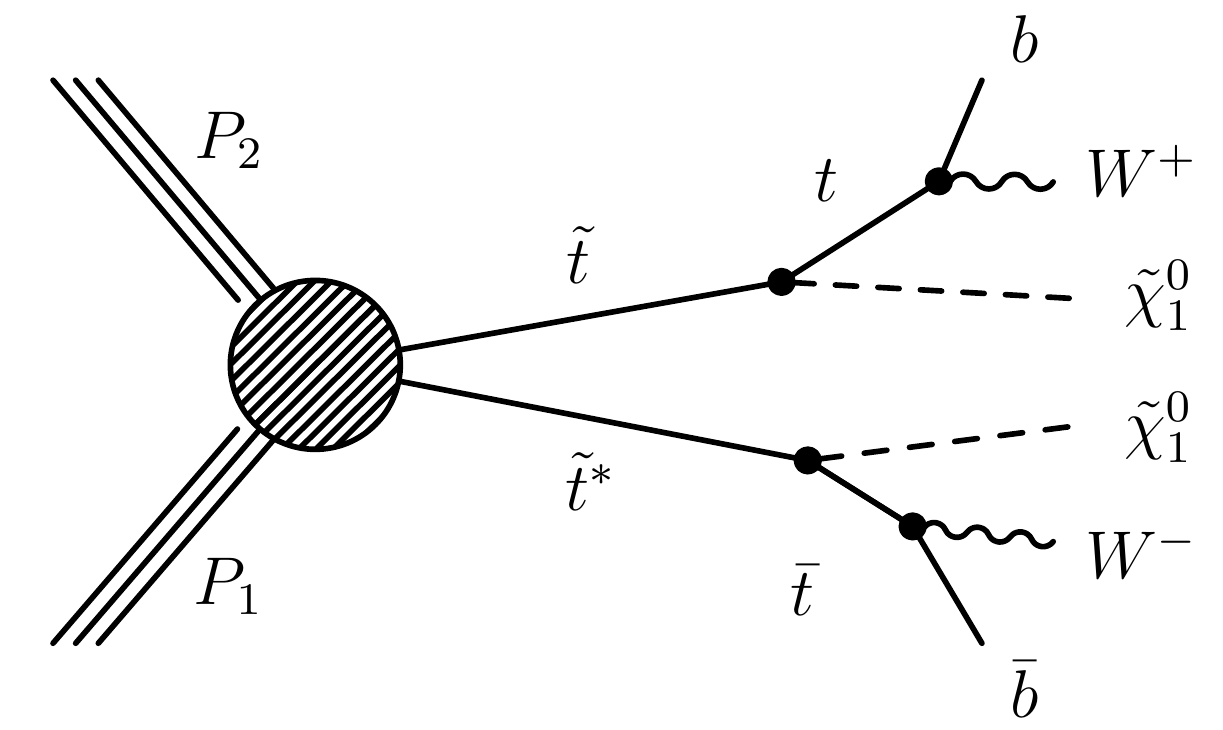}
           \caption{From left to right, diagrams for the simplified models
             $pp\rightarrow\tilde{g}\tilde{g}\rightarrow b\bar{b}b\bar{b}\;\tilde{\chi}_1^0\tilde{\chi}_1^0$, $pp\rightarrow\tilde{g}\tilde{g}\rightarrow
             t\bar{t}t\bar{t}\;\tilde{\chi}_1^0\tilde{\chi}_1^0$, $pp\rightarrow\tilde{g}\tilde{g}\rightarrow
             q\bar{q}q\bar{q}\;\tilde{\chi}_1^0\tilde{\chi}_1^0$, and
             $pp\rightarrow\tilde{t}\bar{\tilde{t}}\rightarrow
             t\bar{t}\;\tilde{\chi}_1^0\tilde{\chi}_1^0$.}
           \label{fig:evtdia}
         \end{figure}

         This search is also highly motivated by models of natural SUSY. In
         these models, the stop, left-handed sbottom, and gluino must have masses near the
         electroweak scale to compensate for large radiative corrections to the
         mass of the higgs boson from SM particles without fine tuning. Although
         there is no way to rigorously define what degree of fine tuning is
         acceptable, a convention among post-Run I natural SUSY models requires
         the masses of the gluino and top squarks to be less than
         2 TeV and 1 TeV, respectively~\cite{papucci,ncraig}. With sufficient data, these models will be probed over
         much of the natural SUSY parameter space in Run II. 
         In $R$-parity conserving SUSY models, in which the LSP is the $\tilde{\chi}_{1}^{0}$, gluinos and squarks are expected to decay to multiple
         hadrons, as well as weakly-interacting SUSY particles that escape detection. To target this signature, we conduct
         our search in a sample of events with multiple jets, large missing transverse momentum ($\MHT$), a large scalar sum of jet transverse momenta ($\HT$), and
         no charged electrons or muons. Diagrams for three of the
         simplified SUSY models we target are shown in Figure \ref{fig:evtdia}. This analysis~\cite{RA2paper} expands and combines strategies from two searches performed by CMS at $\sqrt{s}=8$ TeV~\cite{RA2b8TeV,RA28TeV}.

         \section{Search strategy and event selection}\label{sec:strategy}

         Since SUSY offers an enormous variety of models and final states, we design our search to be inclusive and generic so
         that we maintain sensitivity to a diverse array of new physics
         scenarios. We consider
         events with at least four jets with transverse momenta ($p_T$) of at least 30 GeV, at least 200 GeV of $\MHT$, at least 500 GeV of $\HT$, and no charged leptons. To
         increase sensitivity to different signal topologies, and to help
         characterize a potential excess beyond the measured SM backgrounds, we
         divide our search region into 72 independent search regions defined by
         the jet multiplicity, the $\MHT$,
         the $\HT$, and the b-tagged jet multiplicity of each event. The bins of $\MHT$ and $\HT$ that define the search regions are illustrated in Figure \ref{fig:piecharts} (left). The dominant SM backgrounds for
         the all-hadronic multijet + missing momentum final state include production of top quarks and $W$ and $Z$ bosons in
         association with jets, as well as strongly produced (QCD) multijets. To
         suppress the multijet background, which arises primarily from
         mismeasurement of jet momenta resulting in a fake-$\MHT$ signature,
         we reject events in which any of the four jets of largest $p_T$ is
         aligned with the $\MHT$, specifically, we require
         \begin{eqnarray}
           \Delta\phi(j_i\;,\;\MHT) > 0.5,\;\;\;\; 
           \Delta\phi(j_j\;,\;\MHT) > 0.3
         \end{eqnarray}
         where $i=1,2$ (\underline{i.e.}, the two leading jets) and $j=3,4$.
         Top and
         $W$ events enter the search
         region if a $W$ decays either to an electron or muon that is out of
         kinematic or geometric detector acceptance; if it decays to an
         electron or muon that fails lepton reconstruction, identification, or
         isolation requirements; or if it decays to a tau lepton that decays
         hadronically. To further suppress this background, we reject events
         with an isolated charged track identified by the particle flow
         algorithm~\cite{PF,PF2} as an electron, muon, or charged hadron.

         \section{Background estimation}

         \begin{figure}
           \centering
           \includegraphics[width=0.3\linewidth]{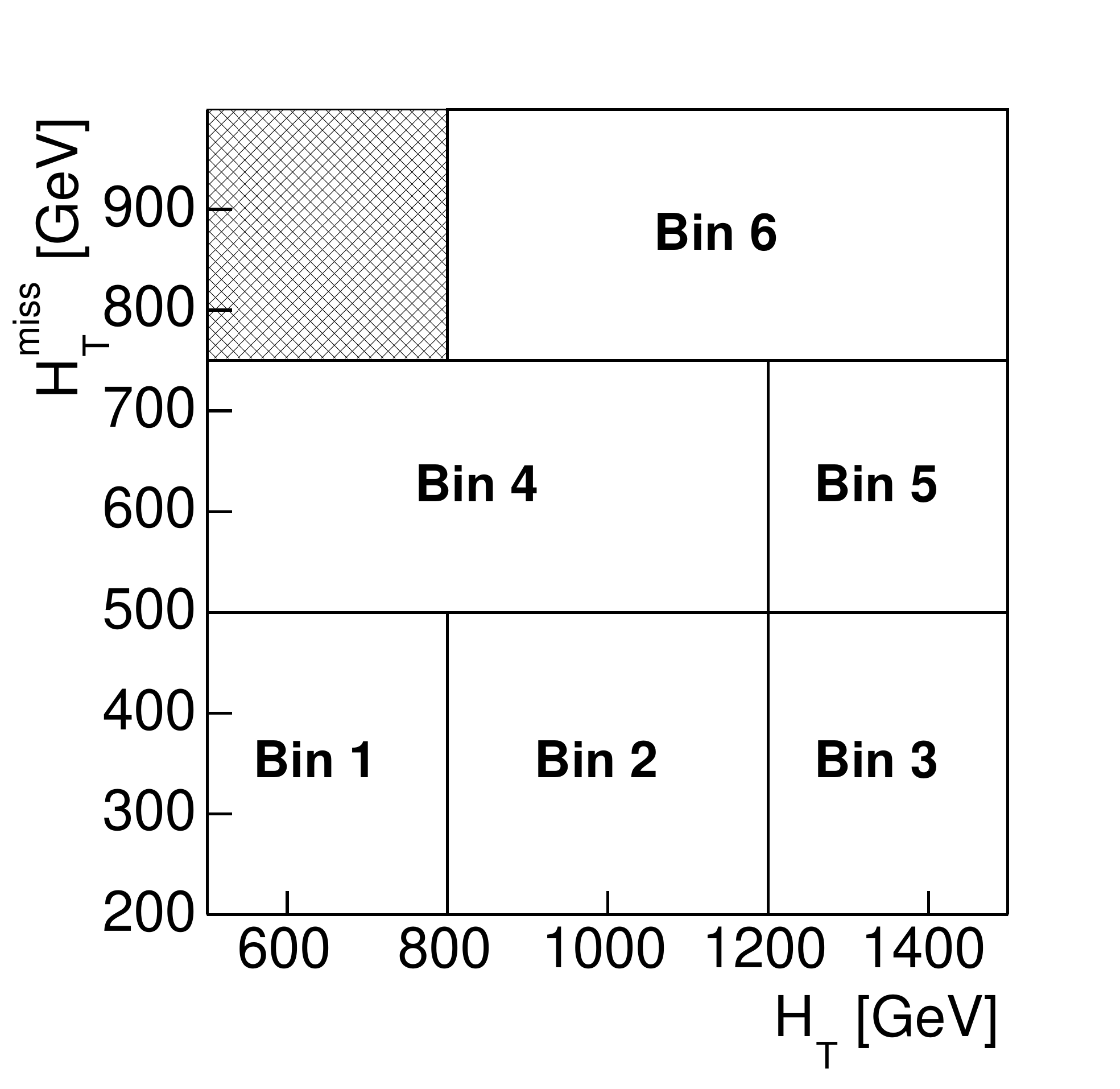}
           \includegraphics[width=0.33\linewidth]{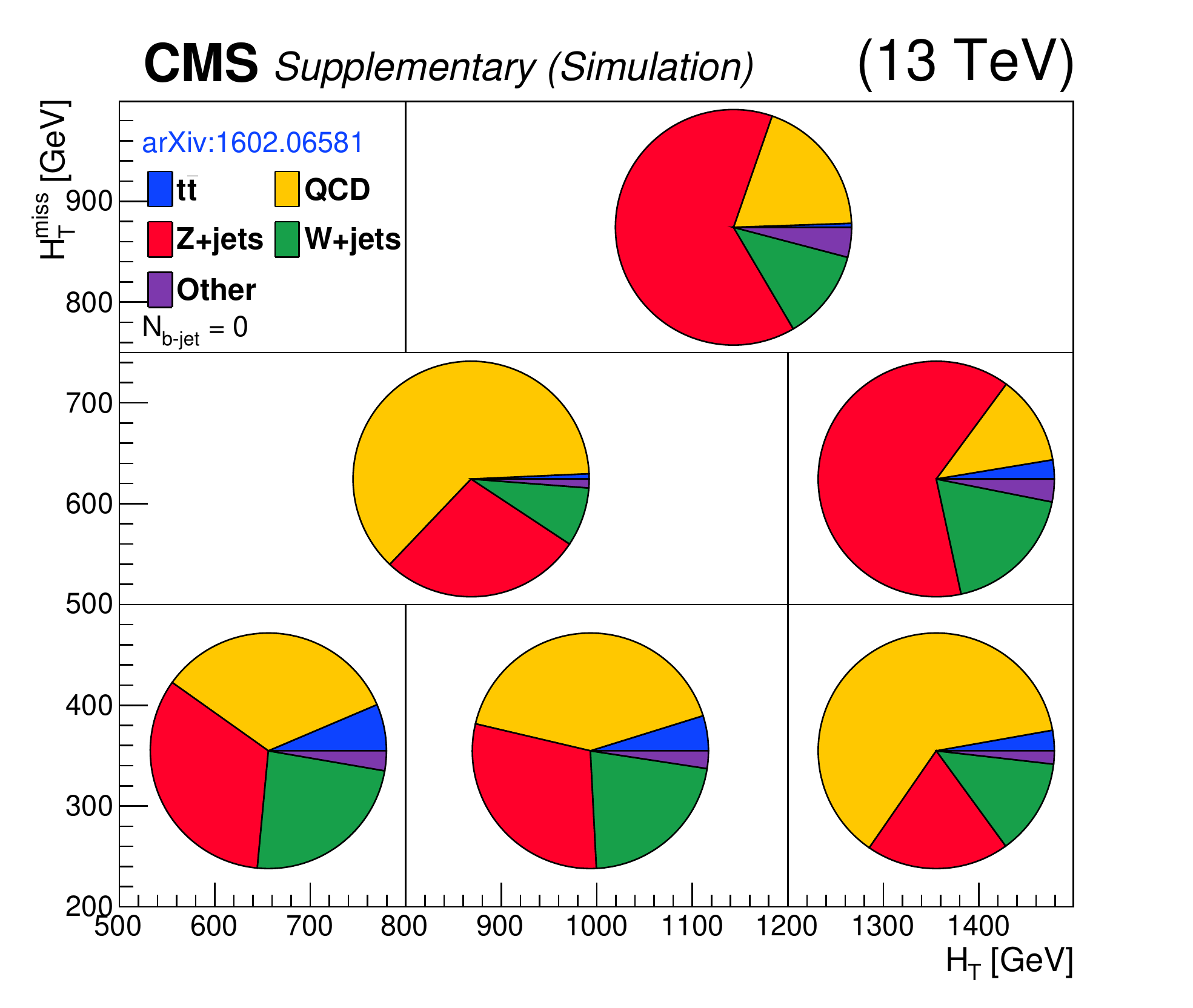}
           \includegraphics[width=0.33\linewidth]{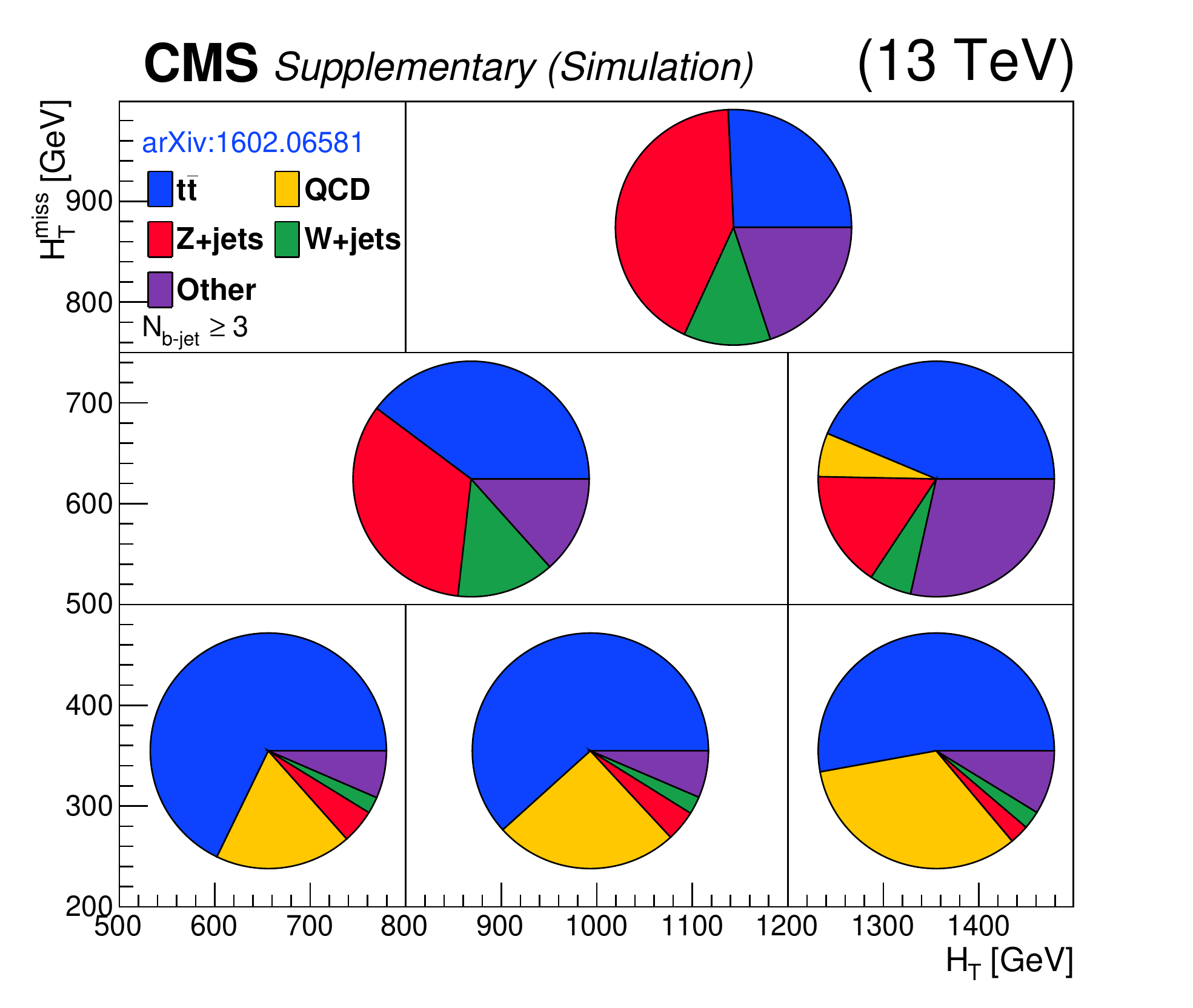}
           \caption{Search bins in the $\MHT$ versus $\HT$ plane (left); background composition in the $\MHT$ versus $\HT$ plane in events with zero b-tagged jets (center); background composition in the $\MHT$ versus $\HT$ plane in events with three or more b-tagged jets (right). The expected
             contribution from each process is obtained from simulation after
             applying the full baseline selection described in Section \ref{sec:strategy}. }
           \label{fig:piecharts}
         \end{figure}

         Following the event selection described in Section \ref{sec:strategy},
         the dominant background in bins with large jet and b-tagged jet
         multiplicity is top quark pair production, while in bins with lower
         multiplicity, the composition is distributed among $W$, $Z$, and QCD
         multijet events. The background composition, as determined in simulation for select search regions, is shown in Figure \ref{fig:piecharts} (center, right). 

         \begin{figure}
           \centering
           \includegraphics[width=0.2425\linewidth]{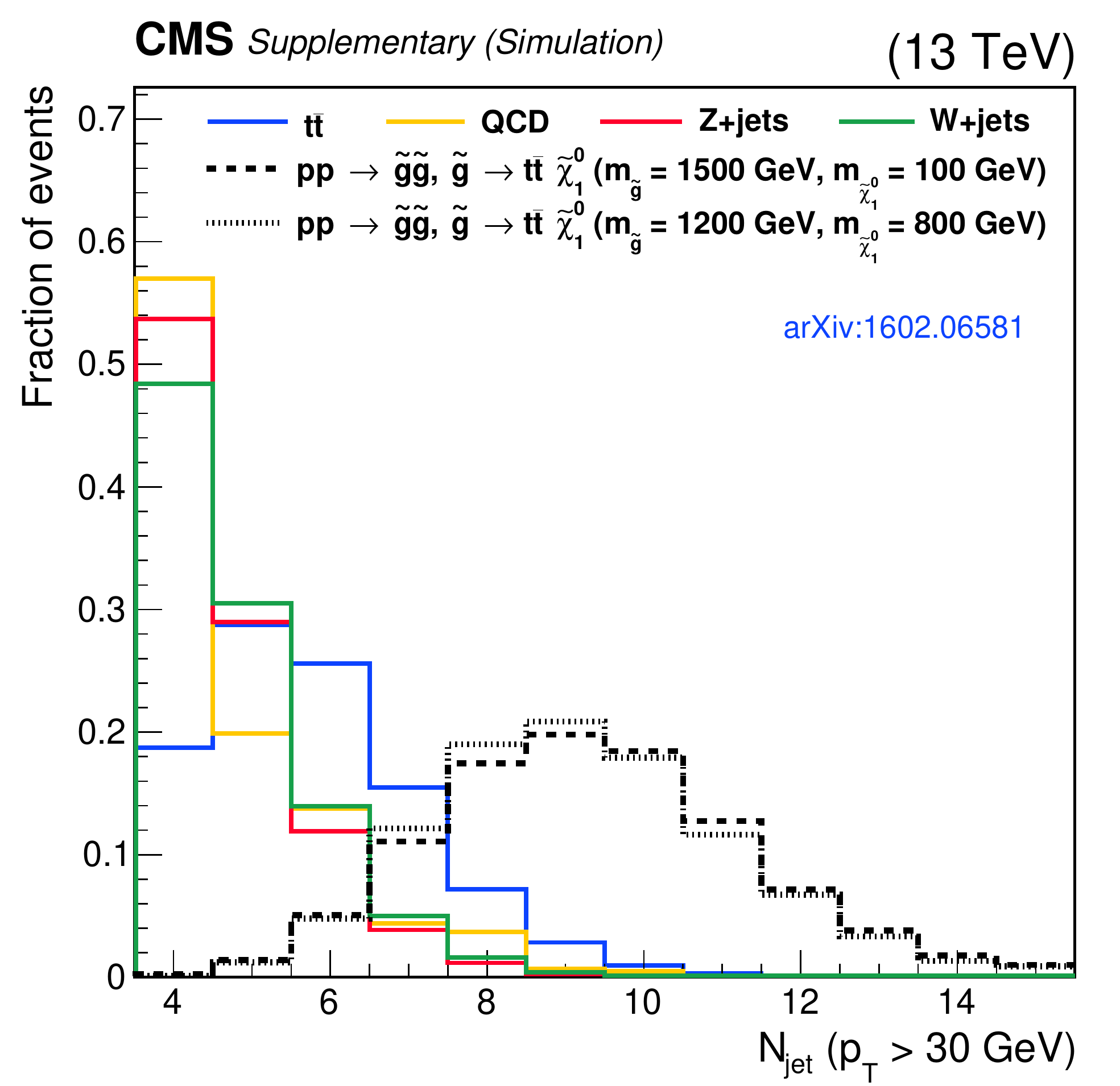}
           \includegraphics[width=0.2425\linewidth]{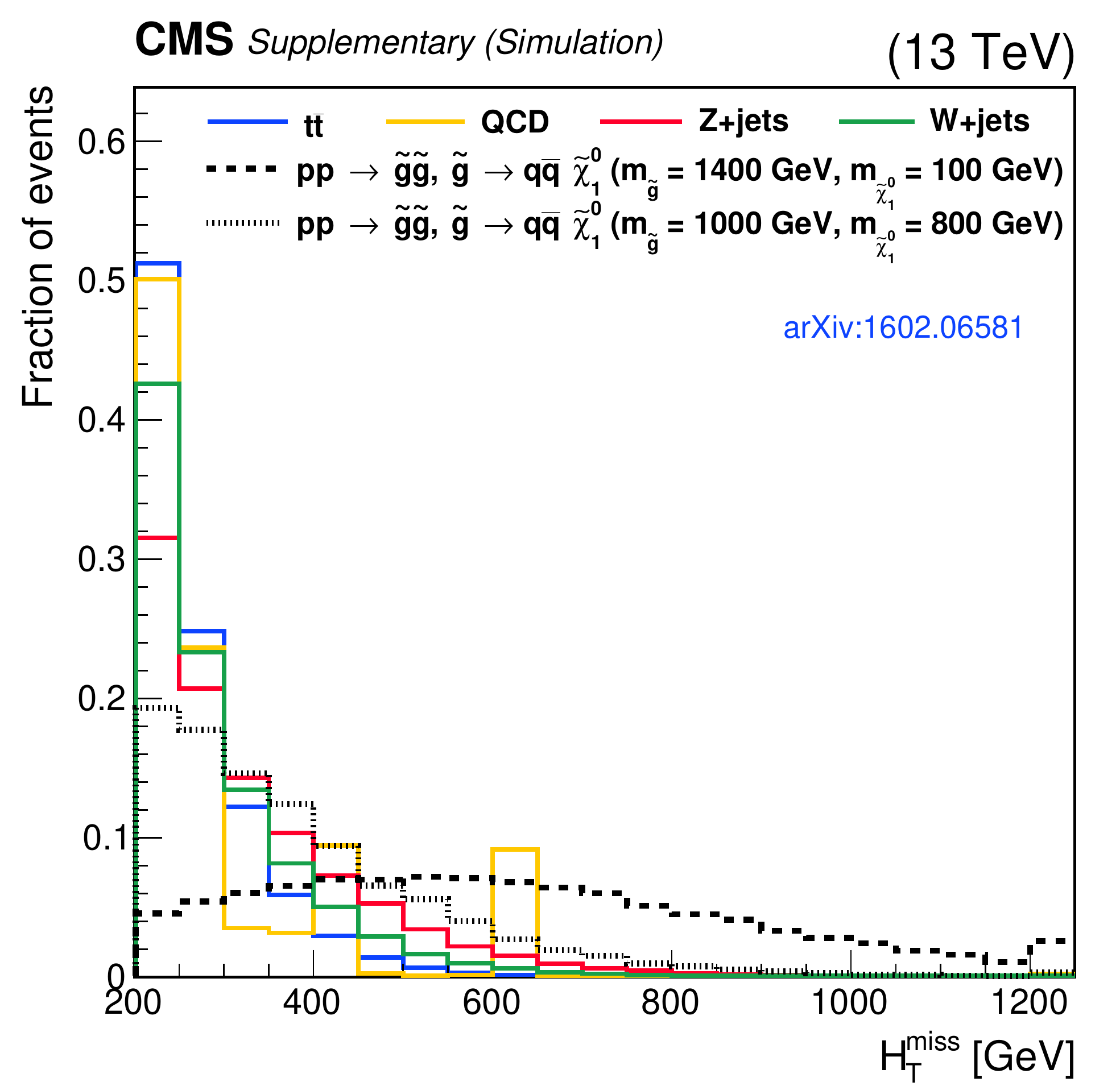}
           \includegraphics[width=0.2425\linewidth]{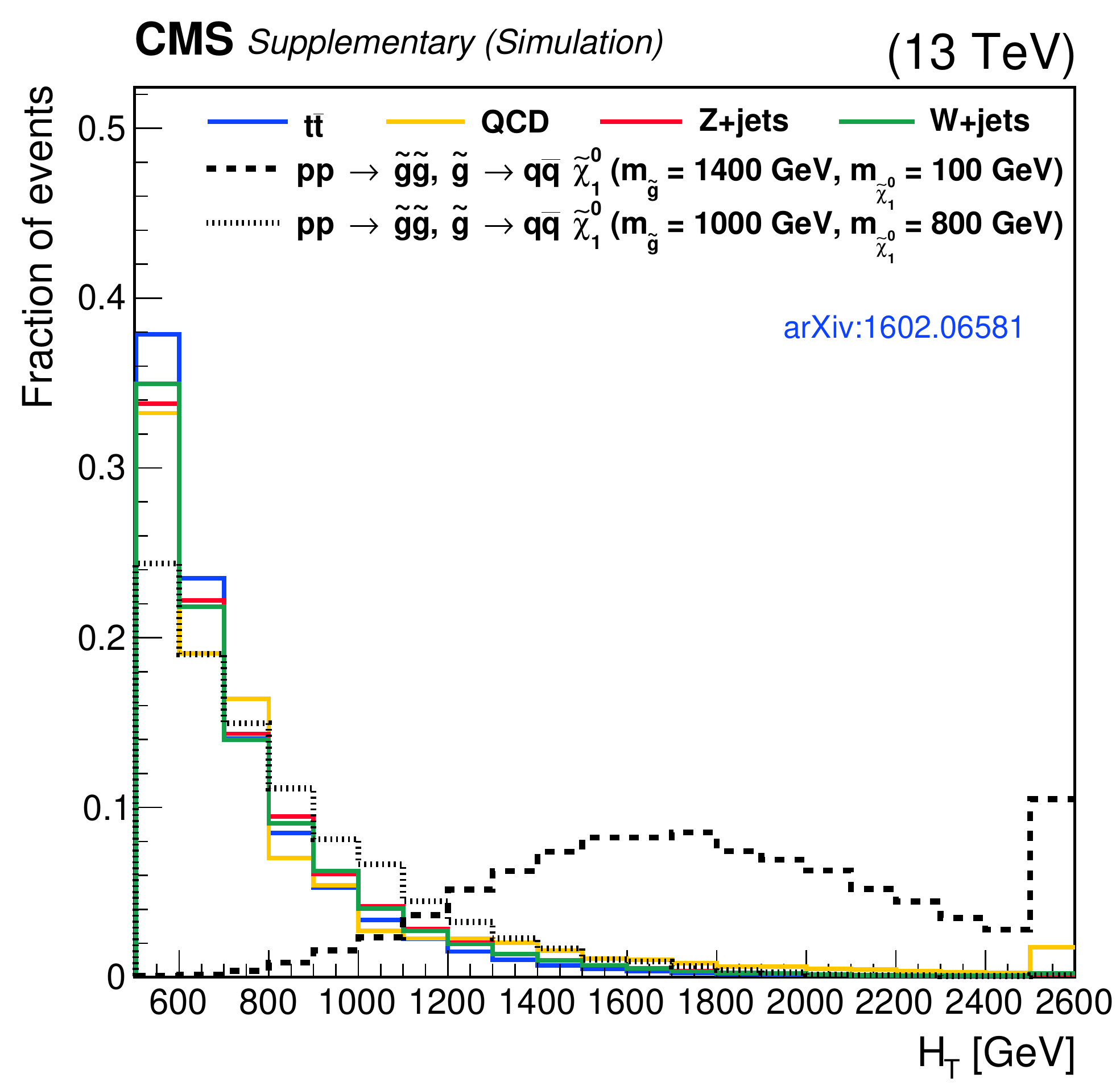}
           \includegraphics[width=0.2425\linewidth]{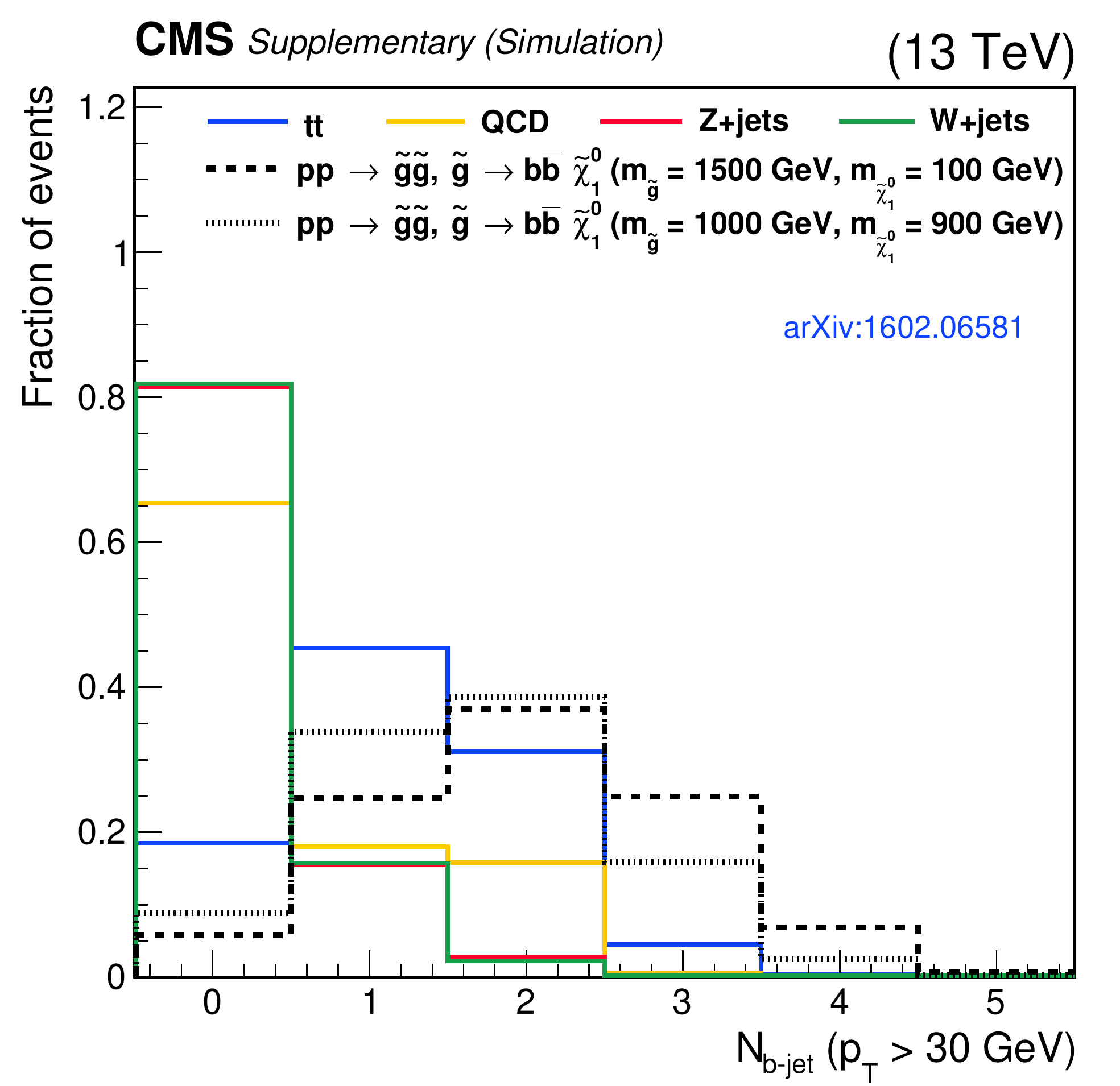}
           \caption{Distributions from simulated event samples showing kinematic shape comparisons of the number of jets, $\MHT$, $\HT$, and the number of b-tagged jets for the main background processes and six representative gluino production signal models. The full baseline selection is applied in each plot.}
           \label{fig:searchshapes}
         \end{figure}

         As the
         cross sections of these backgrounds are generally larger than those of
         our
         target SUSY models by multiple orders of magnitude, many of the search regions
         with highest signal sensitivity lie on extreme tails of 
         kinematic distributions (Figure
         \ref{fig:searchshapes}). The tails of these distributions are difficult
         to model, so we measure their shapes using dedicated techniques and
         control regions (CRs) in data for each background process. The CRs are designed to capture the kinematic properties of the backgrounds in the search region up to differences understood sufficiently well that they can be corrected by simulation. 

         \section{Results}

         \begin{figure}
           \centering
           \includegraphics[width=0.84\linewidth]{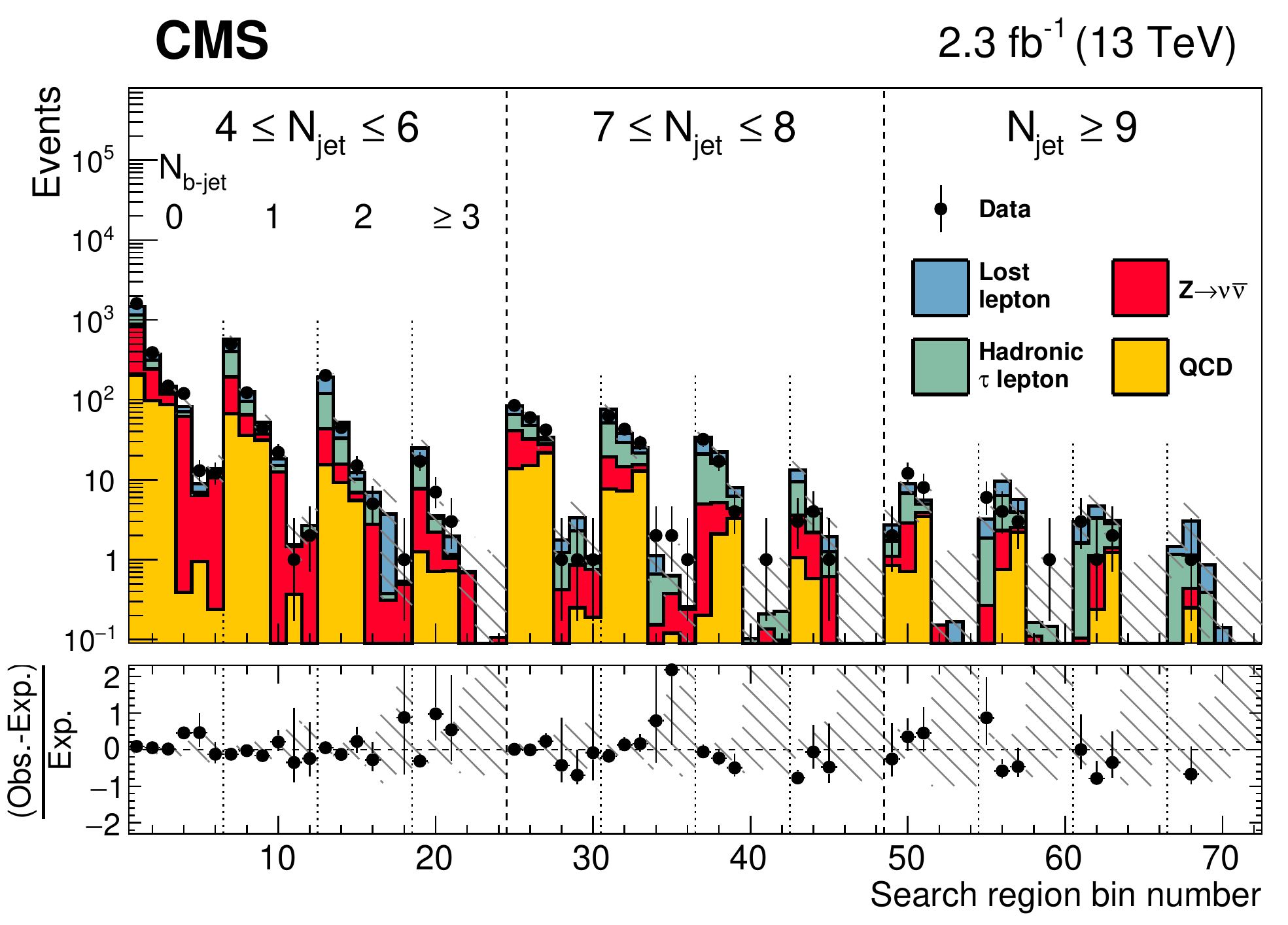}
           \caption{Observed data and summed data-driven background estimation in each of the 72 search regions}
           \label{fig:72bins}
         \end{figure}

         \begin{figure}
           \centering
           \includegraphics[width=0.33\linewidth]{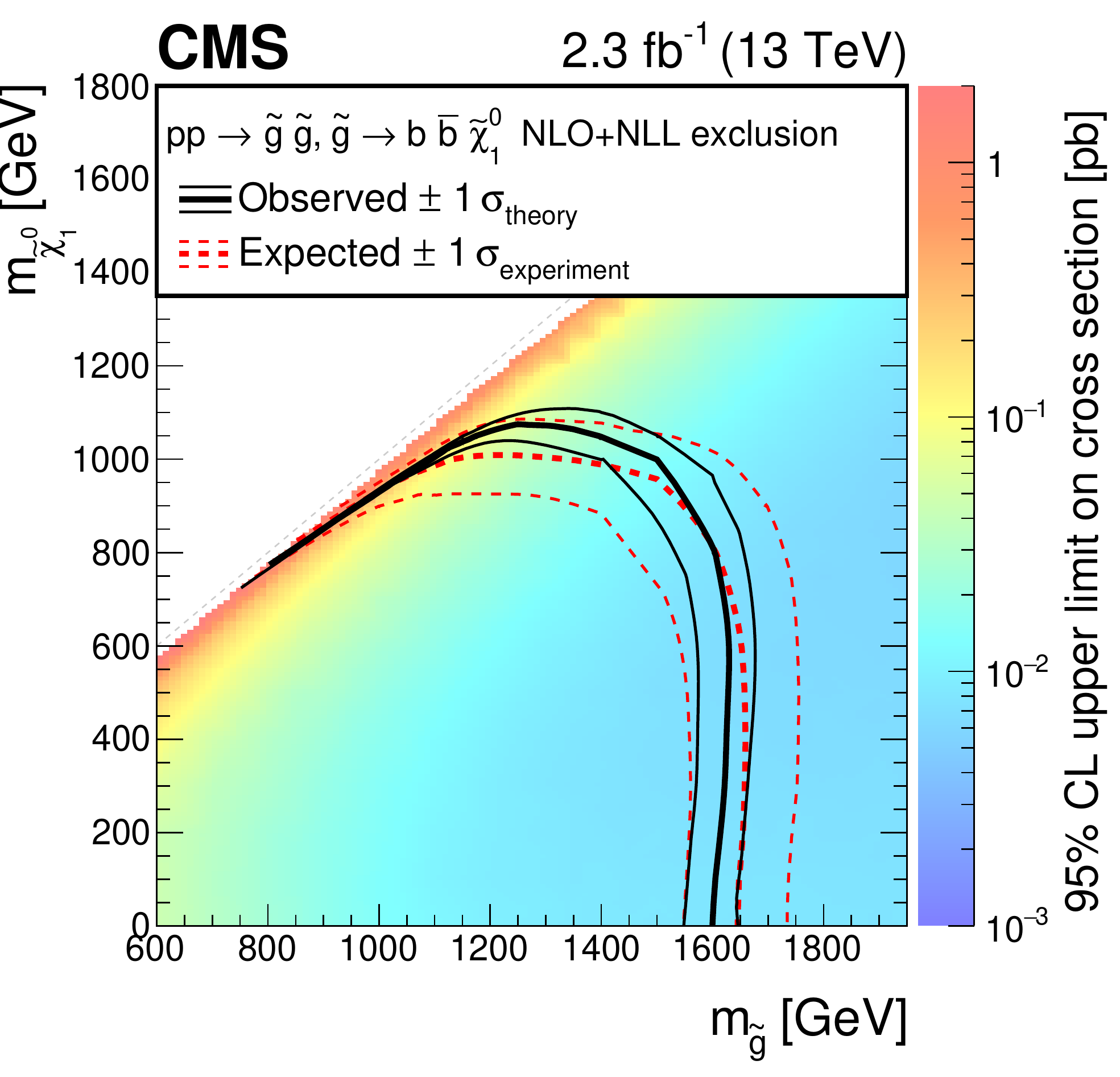}
           \includegraphics[width=0.33\linewidth]{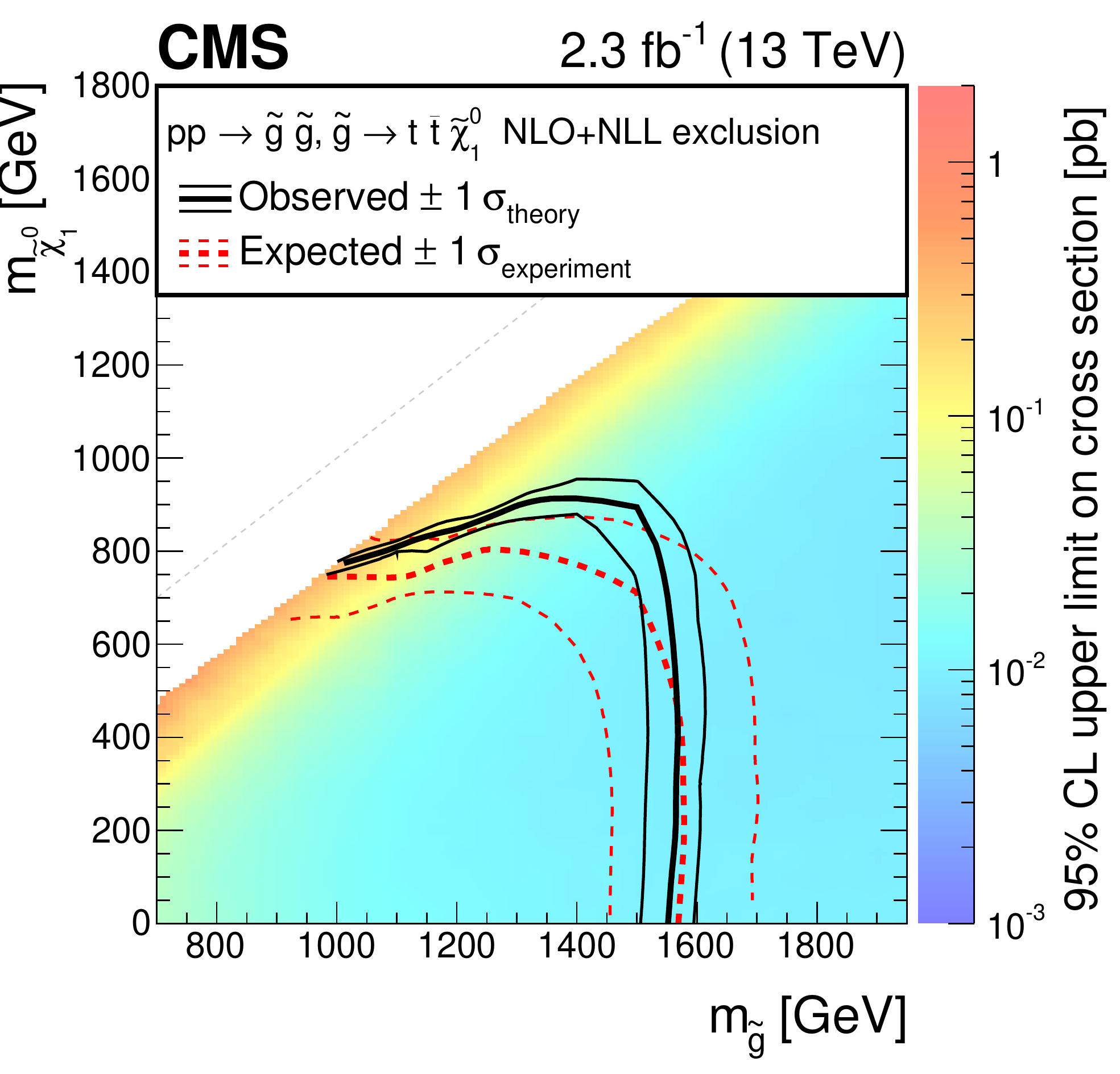}
           \includegraphics[width=0.33\linewidth]{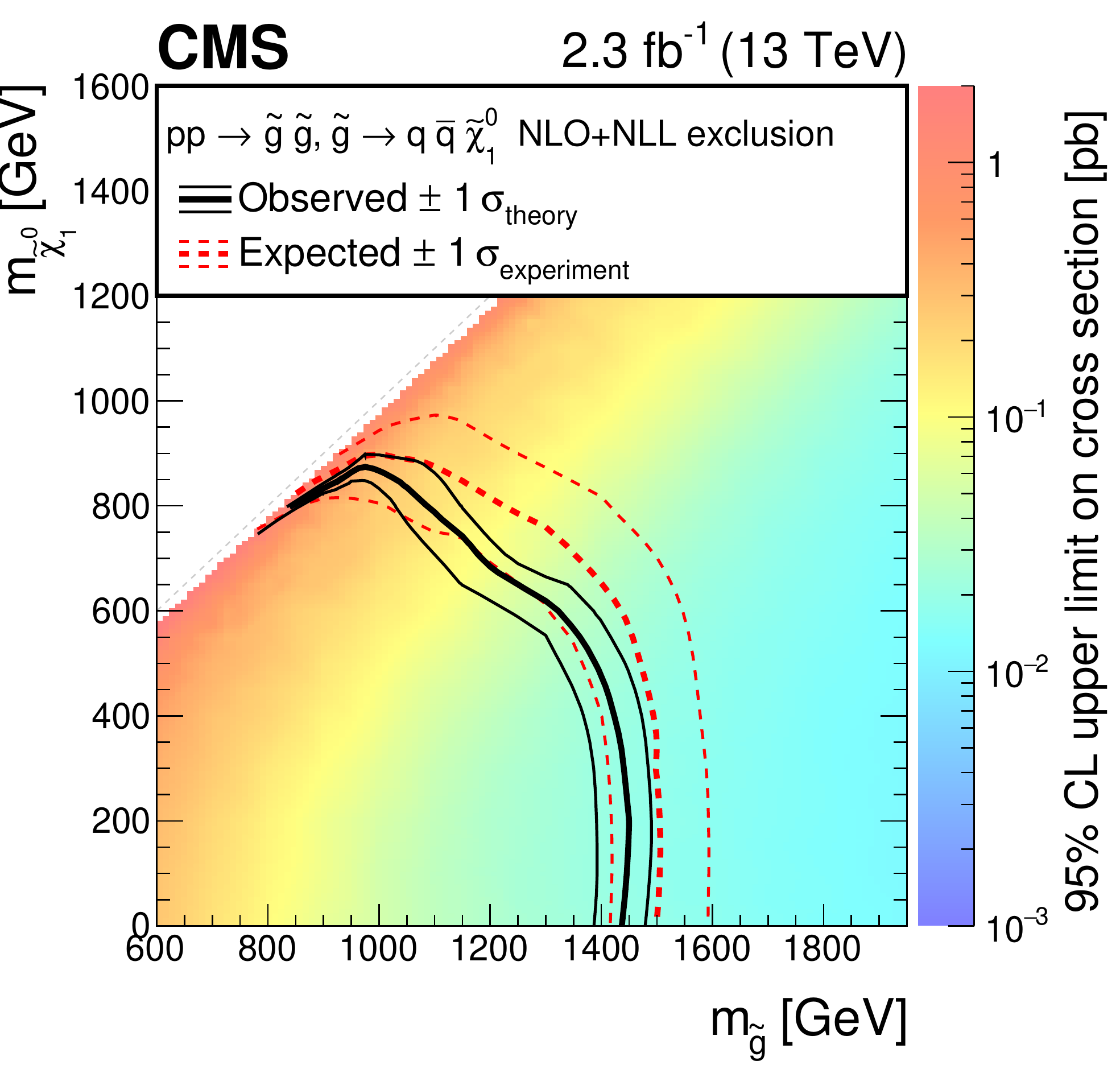}
           \includegraphics[width=0.33\linewidth]{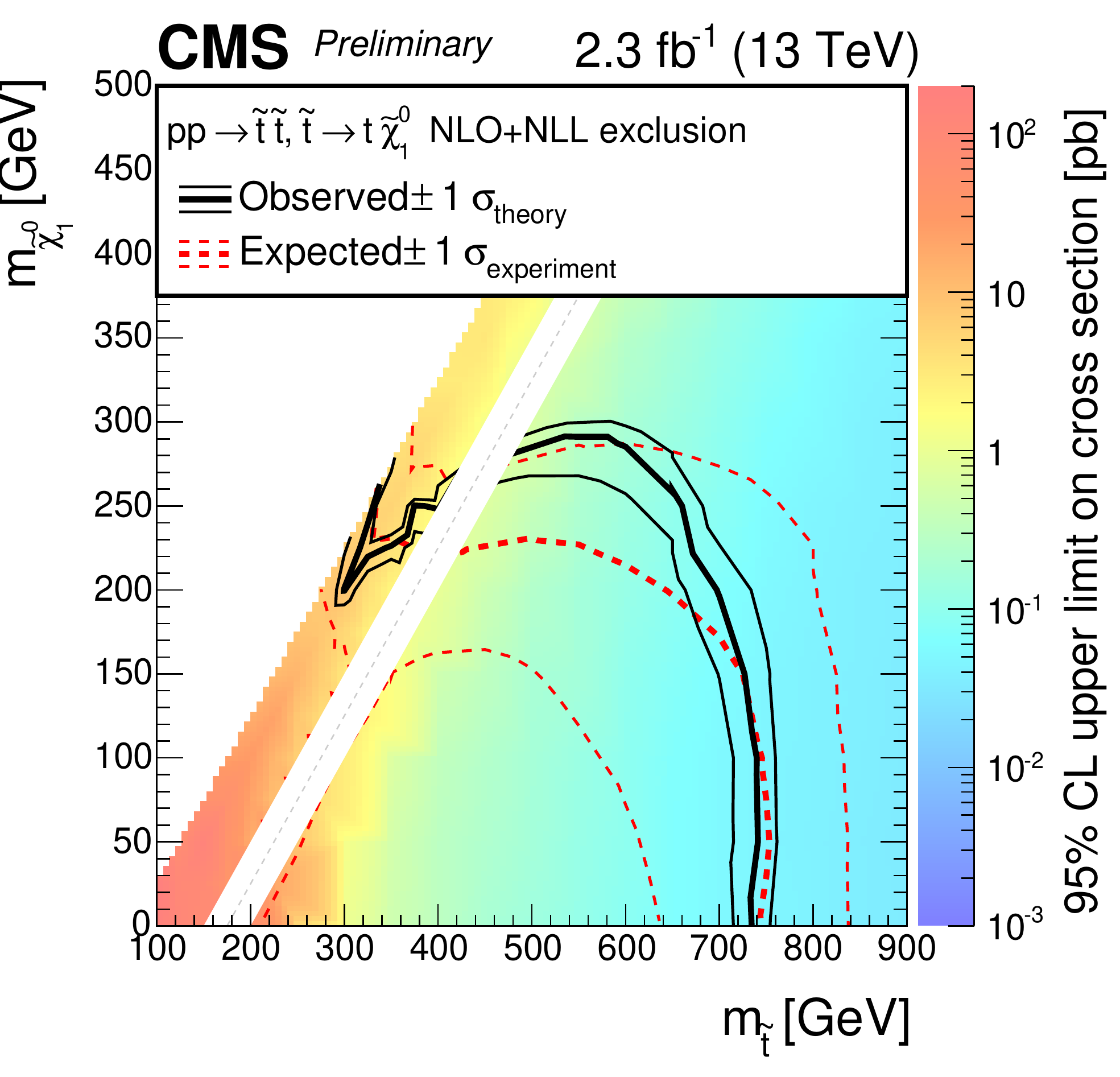}
           \caption{Clockwise from top-left, 95\% CL upper limits on the production cross sections for the simplified models
             $pp\rightarrow\tilde{g}\tilde{g}\rightarrow b\bar{b}b\bar{b}\;\tilde{\chi}_1^0\tilde{\chi}_1^0$, $pp\rightarrow\tilde{g}\tilde{g}\rightarrow
             t\bar{t}t\bar{t}\;\tilde{\chi}_1^0\tilde{\chi}_1^0$,
             $pp\rightarrow\tilde{t}\bar{\tilde{t}}\rightarrow
             t\bar{t}\;\tilde{\chi}_1^0\tilde{\chi}_1^0$, and $pp\rightarrow\tilde{g}\tilde{g}\rightarrow
             q\bar{q}q\bar{q}\;\tilde{\chi}_1^0\tilde{\chi}_1^0$.}
           \label{fig:SMSlimits}
         \end{figure}

         No significant excess is observed beyond the measured SM backgrounds. Figure \ref{fig:72bins} shows the observed data are compared to the summed estimated backgrounds in each of the 72 search regions. The results are interpreted as 95\% CL upper limits on the production cross sections for simplified SUSY models. Three of these interpretations are shown in Figure \ref{fig:SMSlimits}. The strongest mass limits on gluino pair production are obtained for the model in which the gluino decays exclusively to $\tilde{b}\bar{b}$, yielding the striking final state $b\bar{b}b\bar{b}\;\tilde{\chi}_1^0\tilde{\chi}_1^0$. For this model, we can exclude gluino masses below 1600 GeV for light $\tilde{\chi}_1^0$ at the 95\% CL, an improvement of over 200 GeV with respect to the strongest limits set by CMS at $\sqrt{s}=8$ TeV.  For simplified models in which the gluino decays to $\tilde{t}\bar{t}$ and $\tilde{q}\bar{q}$, we exclude gluino masses below 1550 and 1440 GeV, respectively, for light $\tilde{\chi}_1^0$, also significantly extending previous limits. 
         
         \section*{Acknowledgments}

I acknowledge support from the US Department of Energy Office of Science and the Graduate Division of the University of California, Santa Barbara.

       \end{document}